\documentclass[aps,pra,superscriptaddress,amsmath,amssymb,preprintnumbers,floatfix,showpacs,12pt]{revtex4}
\usepackage{amssymb}
\usepackage{epsfig}
\usepackage{subfigure}
\usepackage{graphicx}
\begin{document}
\title{ $SU(2)$- and $SU(1,1)$-squeezing of  interacting radiation modes}

\author{M. Sebawe Abdalla}

\affiliation{Mathematics Department, College of Science, King Saud
University, P.O. Box 2455, Riyadh 11451, Saudi Arabia}

\author{Faisal A. A. El-Orany\footnote{Permanent address: Suez Canal university, Faculty of
Science, Department of mathematics and computer science, Ismailia,
Egypt.}, J. Pe\v{r}ina  } \affiliation{ Joint Laboratory of Optics
of Palack\'y University and Physical Institute of Academy of
Sciences of Czech Republic, 17.~listopadu~50, 772 07~Olomouc,
Czech Republic. }

\begin{abstract}
In this  communication we discuss $SU(1,1)$- and $SU(2)$-squeezing
of an interacting system of radiation modes in a quadratic medium
in the framework of Lie algebra. We show that regardless of which
state being initially  considered, squeezing can be periodically
generated.

\end{abstract}
 \maketitle

\section{Introduction}

The experiments on photon antibunching and sub-Poissonian statistics focused
on the intensity or photon-number fluctuations of electromagnetic field.
Recently, there was a major effort  focused on the fluctuations in the
quadrature amplitudes of the electromagnetic field to produce squeezed light.
This light is indicated by having less noise in one field quadrature
than a coherent state with an excess of noise in
the conjugate quadrature such that the product of canonically conjugate
variances must satisfy the uncertainty relation.
Indeed, this light occupies a wide area in the studies of quantum optics theory since
it has a lot of applications, e.g. in optical
communication networks \cite{yu1}, in interferometric techniques
\cite{ca1}, and in optical waveguide tap \cite{sha}. Moreover,
generation of squeezed light has been observed in many optical
processes, e.g. \cite{lu1,{pi1}}.
Investigation of the squeezing properties of the radiation field is a
central topic in quantum optics and noise squeezing can be measured
by means of homodyne detection.

On the other hand, Lie algebras have been used to investigate the nonclassical
properties of light in quantum optical systems, e.g. quantum mechanical
interferometers \cite{bern}, beam splitters \cite{rich} and linear
directional coupler \cite{coup}, since they can give  powerful and systematic
methods to facilitate such studies \cite{ban}. Among these nonclassical
properties lies $SU(2)$- and $SU(1,1)$-squeezing \cite{eber}.
The authors of \cite{eber} have shown that in the framework of a system of $N$ two-level atoms the squeezing of
angular-momentum [$SU(2)$] fluctuations is exhibited for optical transients
involving the photon echo. Further, the $SU(1,1)$ fluctuations are
established for general two-photon processes involving dynamical variables
different from the creation and annihilation operators.
Also Lie algebra techniques have been applied to problems in nonlinear optics
such as a model of nonabsorbing nonlinear medium (an anharmonic oscillator)
\cite{cher1} or a model consisting of a degenerate parametric amplifier
(nonconserving term) and an anharmonic term \cite{cher2}. For the former
it has been shown that squeezing is eventually revoked and the rate of
revoking grows with
increasing number of photon in the initial state, however, for the latter
 squeezing property is generally revoked by the
nonabsorbing term and  increased by the nonconserving term.
Finally, it is convenient to point out
that the Jaynes-Cummings model composed of a two-level (three-level)
atom interacting with  single mode (two modes) electromagnetic
field has been treated also in terms of Lie algebra \cite{cher3}
(\cite{buz1}). In all these considerations the basic point
is the existence of a set of operators obeying Lie algebra.

The generation of $SU(1,1)$ CS \cite{per1,{baru}}
and  $SU(2)$ CS has been
investigated  for the degenerate \cite{eber} and nondegenerate parametric
amplifiers \cite{eber,{cher4}}, respectively. In this communication we
study $SU(1,1)$- and $SU(2)$-squeezing in terms of these states for
three interacting modes in a nonlinear crystal or in any relevant
device, e.g. nonlinear directional coupler.

This will be done as  follows:
In {\bf section 2} we give a brief overview of the properties of $SU(1,1)$
and $SU(2)$ Lie algebras which will be used in the article. {\bf Section 3}  is devoted
to a discussion of the models as well as  to the solution of the equation
of motions. {\bf Section 4}  discusses $SU(1,1)$-squeezing
and $SU(2)$-squeezing. {\bf Section 5} includes conclusions and remarks.

\section{ Properties of $SU(1,1)$ and $SU(2)$ Lie algebras}
In this section we  review briefly, for future purpose, some properties of
the $SU(1,1)$ and $SU(2)$ Lie algebras as well as we give the notations
of $SU(1,1)$ CS \cite{per1,{baru}} and $SU(2)$ CS \cite{eber}.
We begin by introducing the operators set
 $\{ K_{x} , K_{y},K_{z}\}$ which  satisfy the
commutation relations

${\displaystyle [K_{x}, K_{y}] = i\beta K_{z},    \qquad
[K_{y}, K_{z}] = iK_{x},\qquad [K_{z}, K_{x}] = iK_{y},} \hfill  (1)$

\noindent where $\beta=\pm 1$. When $\beta=-1$ this set becomes the generator
 of $SU(1,1)$ Lie algebra, whereas when $\beta=1$ it becomes the generator
 of $SU(2)$ Lie algebra.
Using the ladder operators, i.e. $K_{+},K_{-}$, we can construct the
operators

${\displaystyle K_{x} = \frac{1}{2}(K_{+}+K_{-}),\qquad \qquad
K_{y} = \frac{1}{2i}(K_{+}-K_{-}),} \hfill (2)$

\noindent satisfying the commutation relation

${\displaystyle [K_{-}, K_{+}] = 2\beta K_{z},   \qquad \qquad
[K_{z}, K_{\pm}] = \pm K_{\pm}. } \hfill (3) $

The discrete representation of the $SU(1,1)$ Lie group
is given by

${\displaystyle K_{z}|m;k\rangle=(m+k)|m;k\rangle, } \hfill $

${\displaystyle
K_{+}|m;k\rangle=[(m+1)(m+2k)]^{\frac{1}{2}}|m+1;k\rangle, }
\hfill $

${\displaystyle
K_{-}|m;k\rangle=[m(m+2k-1)]^{\frac{1}{2}}|m-1;k\rangle, }
\hfill (4) $

\noindent where $K_{-}|0;k\rangle=0 $.
On the other hand, the discrete representation of the $SU(2)$ Lie group
is given by

${\displaystyle K_{z}|m;j\rangle=m|m;j\rangle, } \hfill $

${\displaystyle
K_{+}|m;j\rangle=[(j-m)(j+m+1)]^{\frac{1}{2}}|m+1;j\rangle, }
\hfill $

${\displaystyle
K_{-}|m;j\rangle=[(j+m)(j-m+1)]^{\frac{1}{2}}|m-1;j\rangle, }
\hfill (5) $

\noindent where $K_{-}|-j;j\rangle=K_{+}|-j;j\rangle=0 $.

We examine squeezing against  $SU(1,1)$ CS as well as
$SU(2)$ CS. In fact, there are two types of $SU(1,1)$ CS, the first one is
the PCS \cite{per1}   having the form

${\displaystyle |\xi ;k\rangle =(1-|\xi |^{2})^{k}\sum_{m=0}^{\infty
}\sqrt{\frac{\Gamma (m+2k)}{m!\Gamma (2k)}}\xi ^{m}|m;k\rangle ,} \hfill (6)$

\noindent where $\xi =-\tanh (\frac{\theta}{2})\exp (-i\phi)$, with $
|\xi|\in (0,1),\quad \theta\in (-\infty,\infty),\quad \phi\in (0,2\pi)$,
$\Gamma $ stands for Gamma function and $k$ is called Bargmann index.
For $k=1/4$ and $3/4$ we get even-parity and
odd-parity states, respectively.
This state is a special type of squeezed vacuum state
\cite{eber} which is essentially equivalent to the two-photon coherent state
\cite{yu2}, and it possesses most of the properties of the ordinary
coherent states, such as a completeness relation and a reproducing kernel.
PCS can be realized in the framework of degenerate and
nondegenerate parametric amplifier \cite{cher4}. The second type of
$SU(1,1)$ CS is the
Barut-Girardello  coherent state (BGCS) \cite{baru} determined by

${\displaystyle |z;n\rangle =\sqrt{\frac{|z|^{2n-1}}
{I_{2n-1}(2|z|)}}\sum_{m=0}^{\infty}
\frac{z^{m}}{\sqrt{m!\Gamma (m+2n)}}|m;n\rangle ,} \hfill (7)$

\noindent where $I_{n}(..)$ is the modified Bessel function of order
$n$. Indeed, this state is the eigenstate of $K_{-}$, i.e.
$K_{-}|z;n\rangle =z|z;n\rangle $, and it has  similar properties as
 the Glauber coherent state in the sense that it is not only unsqueezed
state but also  a minimum-uncertainty state.

 $SU(2)$ CS (Bloch state) \cite{eber} is defined by

${\displaystyle |\mu ,j\rangle =\frac{1}
{1+|\mu|^{2}}\sum_{m=-j}^{j}
\sqrt{\frac{(2j)!}{(j-m)!(j+m)!}}\mu^{j+m}|m;j\rangle ,} \hfill (8)$

\noindent where $2j$ is the maximum possible number of photons and
$\mu$ is a complex parameter related to the partition of photons in
the $SU(2)$ CS field modes. This state is squeezed state depending on
the value of $\mu$ and can be generated in a linear
directional coupler in which a pure number state $|2j\rangle$ is launched
into one port of the coupler and the vacuum into other \cite{coup}.

 The following relations will be frequently used in this work \cite{ban}

${\displaystyle
\langle K^{l}_{-}(0)K^{m}_{z}(0)K^{n}_{+}(0)\rangle_{\rm p} =
(1- |\xi|^{2})^{2k} (\frac{\partial}{\partial \xi ^{*}})^{l}
(\frac{\partial}{\partial \xi })^{n}
[k+ |\xi|^{2} \frac{\partial}{\partial (|\xi|^{2}) }]^{m}
\frac{1}{(1- |\xi|^{2})^{2k}}, } \hfill (9a) $

${\displaystyle
\langle K^{l}_{+}(0)K^{m}_{z}(0)K^{n}_{-}(0)\rangle_{\rm b} =
z^{*l}z^{n} \frac{1}{2|z|I_{2n-1}(2|z|)} \left(\frac{x}{2}\frac{\partial}
{\partial x}\right)^{m}xI_{2n-1}(x)|_{x=2|z|}, } \hfill (9b) $

${\displaystyle
\langle K^{l}_{-}(0)K^{m}_{z}(0)K^{n}_{+}(0)\rangle_{\rm u2} =
\frac{1}{(1+|\mu|^{2})^{2j}}
 (\frac{\partial}{\partial \mu ^{*}})^{l}
(\frac{\partial}{\partial \mu })^{n}
[ |\mu|^{2} \frac{\partial}{\partial (|\mu|^{2}) }-j]^{m}
(1+ |\mu|^{2})^{2j}, } \hfill (9c) $

\noindent where the subscripts {\rm p}, {\rm b} and {\rm u2} mean that
the average is performed in terms of PCS, BGCS and $SU(2)$ CS, respectively.

Finally, we  conclude this section by giving the
definitions of $SU(1,1)$- and $SU(2)$-squeezing.
 From (1) we have the following uncertainty relation

\begin{math}
{\displaystyle
\langle (\Delta K_{x} )^{2} \rangle\langle ( \Delta K_{y})^{2}\rangle
 \geq \frac{1}{4} | \langle
K_{z}\rangle |^{2}.} \hfill (10)
\end{math}

\noindent To measure $SU(1,1)$- (or $SU(2)$-) squeezing, it is
appropriate to introduce the function

${\displaystyle
S_{j}=\frac{\langle (\Delta K_{j} )^{2}\rangle- \frac{1}{2} | \langle K_{z} \rangle |
}{%
\frac{1}{2} | \langle K_{z}\rangle |}, \qquad \qquad j=x,y. } \hfill
(11) $

\noindent Maximum $SU(1,1)$- (or $SU(2)$-) squeezing $(100\%)$ is obtained for
$S_{j}=-1$.

\section{ Model description and exact solution}
In this section we consider two types of three radiation modes interacting
by somehow in a nonlinear crystal or in an optical cavity which are
associating with $SU(1,1)$ and $SU(2)$ Lie algebras.

The  lossless effective Hamiltonian of the first type, i.e. associated with
$SU(1,1)$ Lie algebra, has the form

${\displaystyle
\frac{H}{\hbar} =  i \lambda_{1} ( \hat{A}_{1}\hat{A}_{2} -
\hat{A}_{1}^{\dagger}\hat{A}_{2}^{\dagger} ) + i \lambda_{2} (\hat{A}_{1}
\hat{A}_{3} -\hat{A}_{1}^{\dagger}\hat{A}_{3}^{\dagger}) + i \lambda_{3}
(\hat{A}_{3}^{\dagger} \hat{A}_{2} - \hat{A}_{3} \hat{A}_{2}^{\dagger}
), } \hfill (12) $

\noindent where
$\lambda_{j}$ are the coupling constants including the pump amplitude
and proportional to the second-order susceptibility of the
medium $\chi^{(2)}$;
$\omega_{j}$ are the natural  frequencies of oscillation
of the uncoupled modes and {\rm h.c.} is the Hermitian conjugate.
This interaction mixing processes of parametric amplification and
frequency conversion can be established, e.g. by means of
 a bulk nonlinear crystal exhibiting
the second-order nonlinear properties in which three dynamical
modes of frequencies $\omega_1, \omega_2, \omega_3$ are induced
by three beams from lasers of these frequencies.
When pumping this crystal by means of the corresponding
strong coherent pump beams, as indicated in the Hamiltonian,
we can approximately fulfil the phase-matching conditions for the
corresponding processes, in particular if the frequencies are close
each other (biaxial crystals may be helpful in such an arrangement).
Also a possible use of quasi-phase matching may help in the realization,
which is, however, more difficult technologically \cite{real}.
Another  possibility to realize such interaction
is a nonlinear symmetric directional coupler
composed of two  nonlinear waveguides operating by
nondegenerate parametric amplification
where  the interaction
between  two waveguides can be established through the evanescent waves.
More details about the quantum properties of the Hamiltonian  (12)
can be found in \cite{abd1}.
Now if we set

${\displaystyle L_{x} =  i (\hat{A}_{1} \hat{A}_{2} -
\hat{A}_{1}^{\dagger}\hat{A}_{2}^{\dagger}), } \hfill  $

${\displaystyle L_{y} =  i ( \hat{A}_{1} \hat{A}_{3} - \hat{A}
_{1}^{\dagger} \hat{A}_{3}^{\dagger}), } \hfill  $

${\displaystyle L_{z} = i ( \hat{A}_{3}^{\dagger} \hat{A}_{2} -
\hat{A}_{3}  \hat{A}_{2}^{\dagger}), } \hfill (13) $

\noindent one can easily verify that this set of operators satisfy
the commutation rules (1) with $\beta=-1$, i.e. this model is
associated with $SU(1,1)$ Lie algebra.

The second type of Hamiltonian which  associates with
$SU(2)$ Lie algebra has the form

${\displaystyle
\frac{H}{\hbar} =   i \lambda^{'}_{1}
(\hat{A}_{3}^{\dagger}\hat{A}_{2}- \hat{A}_{3} \hat{A}_{2}^{\dagger} )
+ i \lambda^{'}_{2}(
\hat{A}_{1}^{\dagger}\hat{A}_{3}-\hat{A}_{1} \hat{A}_{3}^{\dagger} )
+ i \lambda^{'}_{3}
(\hat{A}^{\dagger}_{1} \hat{A}_{2} - \hat{A}_{1} \hat{A}_{2}^{\dagger} )
, } \hfill (14) $

\noindent where all the notations have the same meaning as before; this
interaction is mixing three processes of frequency conversion.
Analogously if one takes the terms involving $\lambda^{'}_{1},
\lambda^{'}_{2}$ and $\lambda^{'}_{3}$ by $ L^{'}_{x}, L^{'}_{y}$ and
$L^{'}_{z}$, respectively, it is easy to prove that these operators satisfy
the commutation rules (1) with $\beta=1$.
For completeness, it would be convenient to mention
that  pair creation and annihilation operators
$\hat{A}_{j}\hat{A}_{k}$ and $\hat{A}^{\dagger}_{j}\hat{A}^{\dagger}_{k}$
$(j\neq k)$ of the two-mode field form elements of the $SU(1,1)$ Lie
group; on the other hand operators
 $\hat{A}^{\dagger}_{j}\hat{A}_{k}$ and
$\hat{A}_{j}\hat{A}_{k}^{\dagger}$ form
 elements of the $SU(2)$ Lie group, e.g.  in the lossless beam splitter
 \cite{rich}.  So one can note that (12) includes three terms, two of them
represent
$SU(1,1)$ Hamiltonian (parametric amplifiers, $L_{x},L_{y}$)  and
the third one
forms $SU(2)$ Hamiltonian ($L_{z}$). Hamiltonian in (14) is formally
a sum of three $SU(2)$ Hamiltonians. We assume that the used optical crystal is pumped
simultaneously in two different regimes by corresponding laser beams.

Now, in this paper we treat the
systems (12) and (14) by  unified model that exploits their
underlying Lie algebra similarity. The unified model is

${\displaystyle \frac{H}{\hbar} = \alpha_{1} K_{x} + \alpha_{2}
K_{y} +\alpha_{3} K_{z},} \hfill (15) $

\noindent where  $\alpha_{j},\quad j=1,2,3$  are parameters
specializing which model is considered. In other words,
the Lie algebras results for either
 the models (12) or (14) can be recovered from our general formula (15) by
taking $\beta=+1$ or $-1$ for $K_{j}=L_{j}$ or
$K_{j}=L^{'}_{j}, j=x,y,z$, respectively, and specializing the constants
$\alpha_{j}$ to the particular values that they  have in
 the corresponding models $(\lambda_{j}$ or $\lambda^{'}_{j},\quad
 j=1,2,3)$. Now the energy of the system is
proportional to Lie algebra generators. It would be of interest to mention
that a similar model of (15) has been considered in \cite{arv} for
semiclassical Dicke model and the ideal parametric amplifier, however
the treatments there have been given in the framework of pseudospin
vector and/or pseudotensor and consequently simple geometrical arguments have
been performed to explain the phenomena. Indeed, the model (15) is quite
general for any operator system can fulfill the
$SU(1,1)$ or $SU(2)$ Lie
algebra rules.

To discuss the dynamical behaviour of the model we may solve the Heisenberg
 equations of motion for the Hamiltonian (15) which
are

${\displaystyle \frac{dK_{x}}{dt} = -\alpha_{3} K_{y} + \beta\alpha_{2}
K_{z}, }\hfill  $

${\displaystyle \frac{dK_{y}}{dt} = \alpha_{3} K_{x} - \beta\alpha_{1}
K_{z}, }\hfill  $

${\displaystyle
\frac{dK_{z}}{dt} = -\alpha_{2} K_{x} + \alpha_{1} K_{y}. } \hfill (16) $

\noindent The matrix representation of the solutions of these
 equations is

${\displaystyle
\left[
\begin{array}{l}
K_{x}(t) \\
K_{y} (t) \\
K_{z} (t)
\end{array}
\right] = \left[
\begin{array}{lll}
R_{1} (t,\beta) & J^{(-)} (t,\beta) & \beta S^{(+)}(t) \\
J^{(+)}(t,\beta) & R_{2} (t,\beta) & \beta V^{(-)} (t) \\
S^{(-)}(t) &  V^{(+)}(t) & R_{3}(t,\beta=1)
\end{array}
\right] \left[
\begin{array}{l}
K_{x} (0) \\
K_{y} (0) \\
K_{z} (0)
\end{array}
\right], } \hfill (17) $

\noindent where

${\displaystyle R_{j} (t,\beta) =  \cos (gt) + 2 \frac{\beta
\alpha_{j}^{2}}{g^{2}} \sin^{2} (\frac{gt}{2}), \quad j =
1,2,3,}
\hfill  $

${\displaystyle J^{(\pm)}(t,\beta) =  2 \frac{\beta\alpha_{1}\alpha_{2}}{g^{2}}
\sin^{2} (\frac{gt}{2}) \pm \frac{ \alpha_{3}}{g} \sin (g t ), }
\hfill $

${\displaystyle S^{(\pm)}(t) = 2
\frac{\alpha_{1}\alpha_{3}}{g^{2}}
\sin^{2} (\frac{gt}{2}) \pm \frac{ \alpha_{2}}{g} \sin (g t), }
\hfill $

${\displaystyle V^{(\pm)}(t) = 2 \frac{\alpha_{2}\alpha_{3}}{g^{2}}
\sin^{2} (\frac{gt}{2}) \pm \frac{ \alpha_{1}}{g} \sin (g t),}
\hfill (18) $

\noindent and $g = (\alpha_{3}^{2} + \beta\alpha^{2}_{1}
+ \beta\alpha^{2}_{2})^{\frac{1}{2}}$.
 It is easy to check that the commutation relations (1) are
still  valid for solutions (17). Moreover, this solution is periodic
with period $4\pi/g$, i.e. $K_{j}(t+\frac{4n\pi}{g})=K_{j}(t), \quad
n=0,1,2,..$ provided that $g$ is real.
It is reasonable mentioning that one can alternatively work in
the Schr\"{o}dinger picture where the operators
remain unchanged but the state vector of the model  becomes
time-dependent, i.e.
 $|\psi (t)\rangle=
\exp(-itH) |\psi (0)\rangle$ where $|\psi (0)\rangle$ is the initial state
of the system. Then using the disentanglement theorem of
$SU(1,1)$ or $SU(2)$ Lie algebra \cite{ban} the problem can be treated
in an algebric way.

Based on the results of the present section together with those of the
2nd section we discuss the $SU(1,1)$- and $SU(2)$-squeezing in the
following section.
\begin{figure}[h]%
  \centering
  \subfigure[]{\includegraphics[width=6cm]{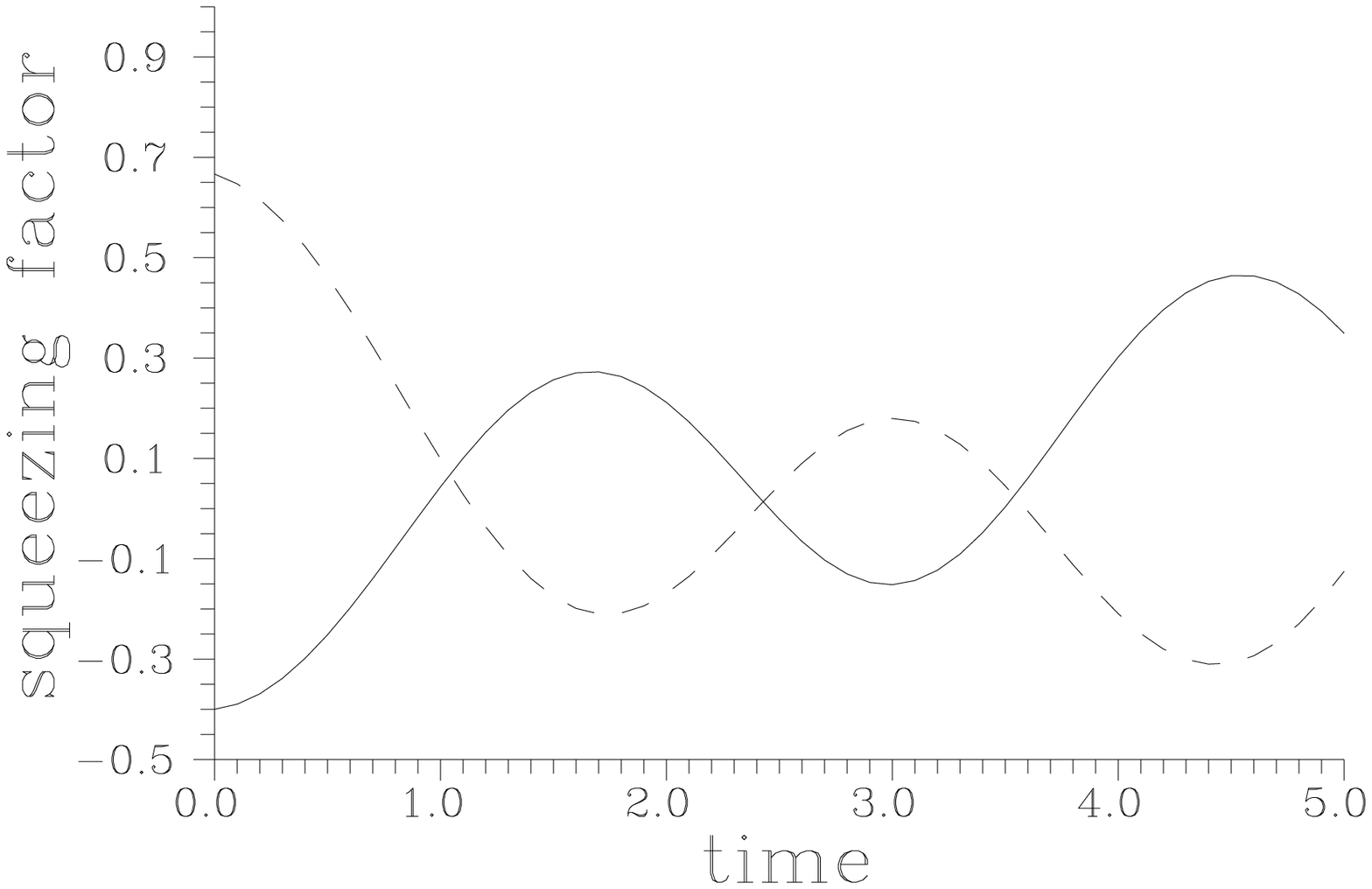}}
 \subfigure[]{\includegraphics[width=6cm]{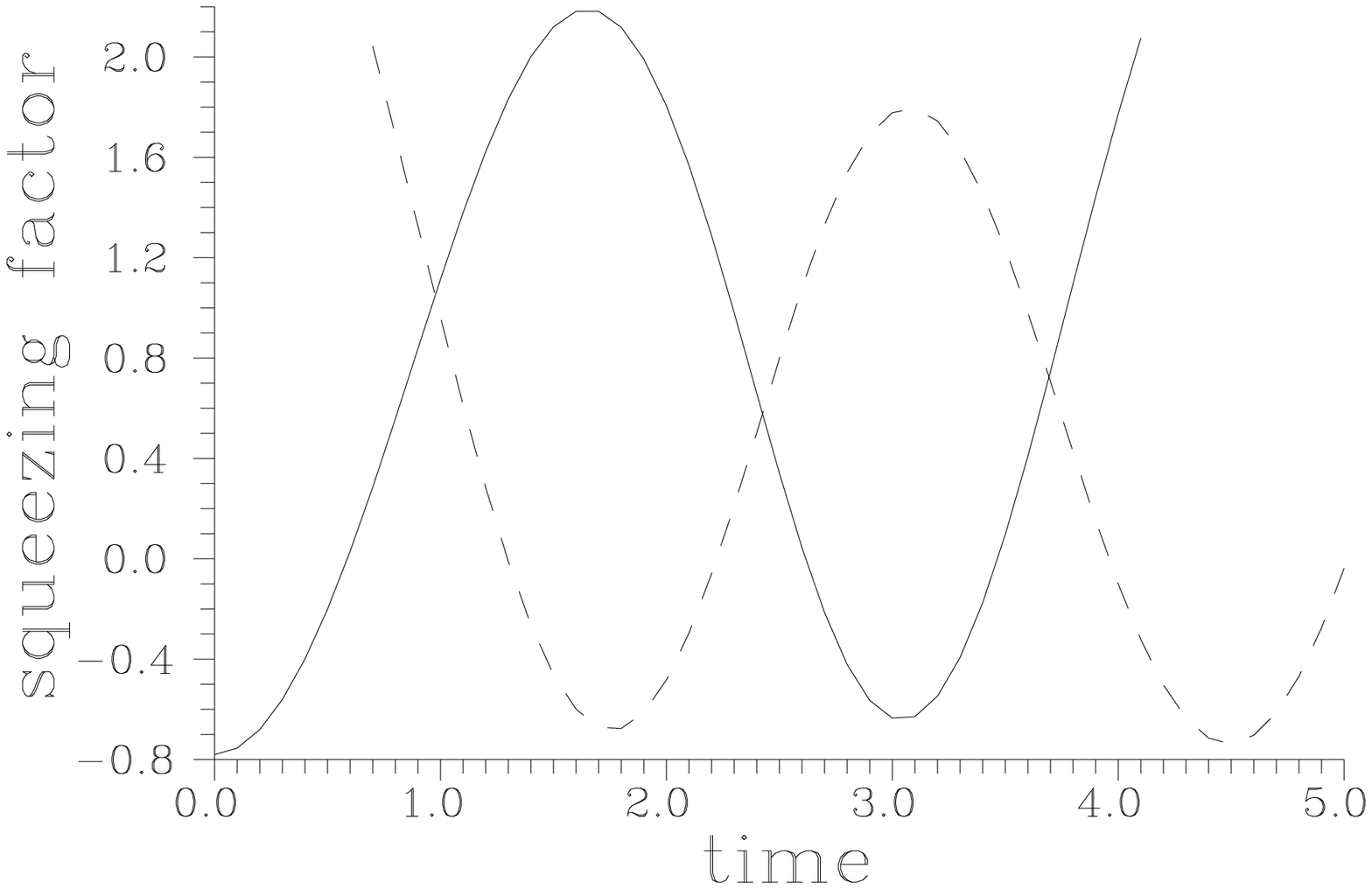}}
 \subfigure[]{\includegraphics[width=6cm]{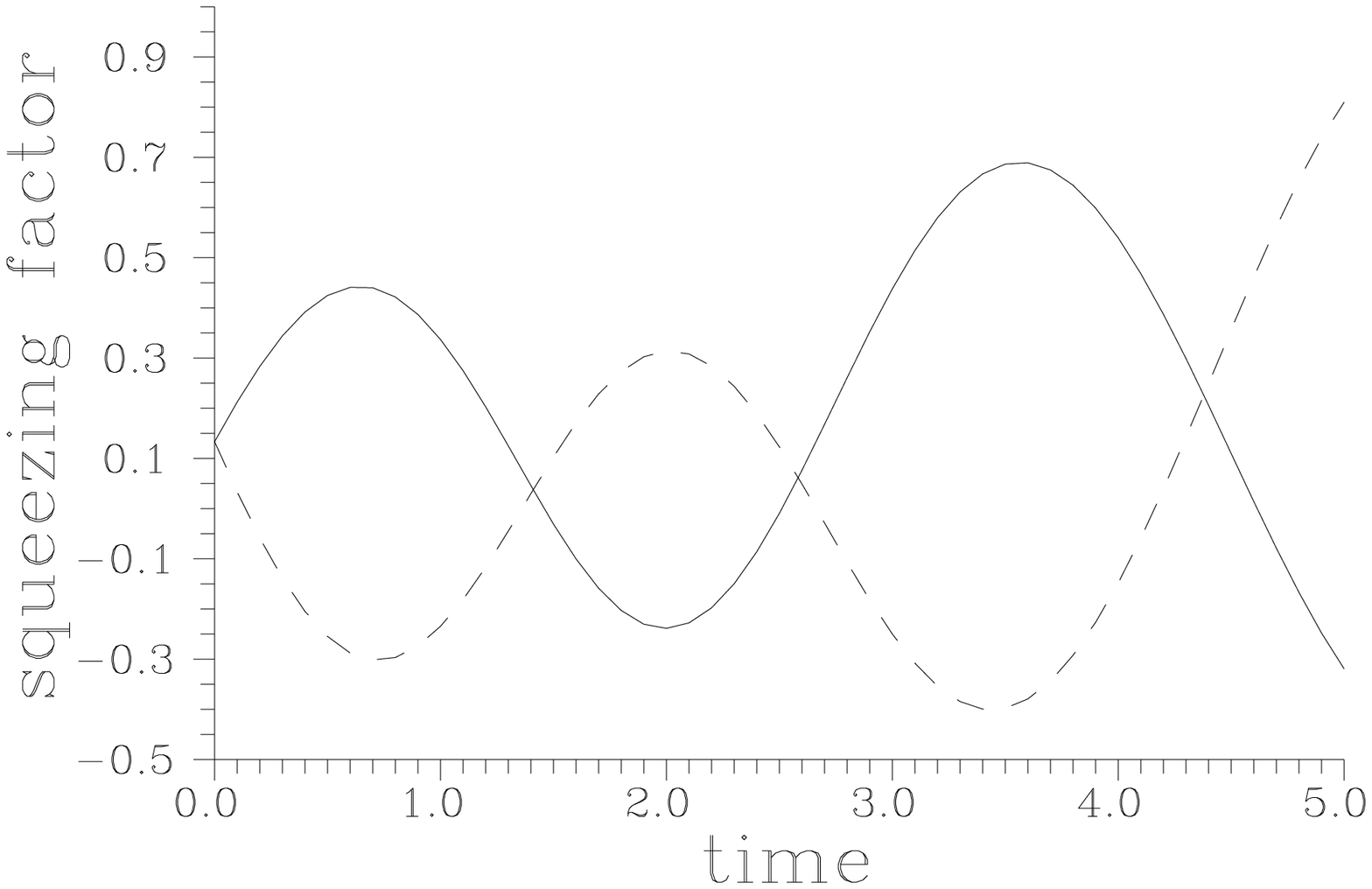}}
    \caption{
Squeezing factor $S_{j}(t)$ of PCS against time $t$ for
$(\lambda_{1},\lambda _{2},\lambda _{3})=(0.1,0.25,1)$ and for a)
$ (\phi,|\xi |)=(\frac{\pi }{2},0.5)$; b) $ (\phi,|\xi
|)=(\frac{\pi }{2},0.8)$; c) $ (\phi,|\xi |)=(\frac{\pi
}{4},0.5)$. In these figures first quadrature is always
represented by the solid curve, whereas second quadrature is
represented by the dashed curve. }
  \label{fig1}
\end{figure}

\section {$SU(1,1)$- and $SU(2)$-squeezing}
First, we consider the $SU(1,1)$-squeezing and investigate fluctuations
in terms of  PCS and BGCS. For this purpose,
the relations (6), (7), (9a-b) and (17) should be used.

After some calculations  the quadrature variances
$\langle (\Delta K_{x}(t))^{2}\rangle_{\rm p}$
 and $\langle (\Delta K_{y}(t))^{2}\rangle_{\rm p} $ as well as $\langle
 K_{z}(t)\rangle_{\rm p}$  for PCS can be written in the following
forms

${\displaystyle \langle (\Delta K_{x}(t))^{2}\rangle_{\rm p} = 2k \left\{
|f(t,-1)|^{2}+\frac{[S^{(+)}(t)-\xi^{*}f(t,-1)-\xi
f^{*}(t,-1)]^{2}}{(1-|\xi|^{2})^{2}}\right.} \hfill  $

${\displaystyle
+\left.\frac{S^{(+)}(t)[\xi^{*}f(t,-1)+\xi f^{*}(t,-1)-S^{(+)}(t)]}
{(1-|\xi|^{2})}\right\}}, \hfill (19a) $

$\hfill $

${\displaystyle \langle (\Delta K_{y}(t))^{2}\rangle_{\rm p} = 2k \left\{
|g(t,-1)|^{2}+\frac{[V^{(-)}(t)-\xi^{*}g(t,-1)-\xi
g^{*}(t,-1)]^{2}}{(1-|\xi|^{2})^{2}}\right.} \hfill  $

${\displaystyle
+\left.\frac{V^{(-)}(t)[\xi^{*}g(t,-1)+\xi g^{*}(t,-1)-V^{(-)}(t)]}
{(1-|\xi|^{2})}\right\}}, \hfill (19b) $

\noindent and

${\displaystyle
\langle K_{z}(t)\rangle_{\rm p} = \frac{k}{(1- |\xi|^{2})} \left\{
(1+ |\xi|^{2})
R_{3}(t,1) + 2[\xi^{*}h(t)+\xi h^{*}(t)]\right\}, } \hfill (19c) $

\noindent where  we have used the following
abbreviations

${\displaystyle
f(t,\beta)=\frac{1}{2}
[ R_{1}(t,\beta)-iJ^{(-)}(t,\beta)], \quad \quad
g(t,\beta)=\frac{1}{2}
[J^{(+)}(t,\beta) -iR_{2}(t,\beta)], } \hfill  $

${\displaystyle
h(t)=\frac{1}{2}[S^{(-)}(t) -iV^{(+)}(t)].} \hfill (20) $

\noindent Of course, $\beta=-1$ in the present case.

It is easy to check that relations (19) reduce to those of
\cite{ban,{cher4}} at $t=0$.
From (11) and (19) it is evident that the fluctuations are independent of
the value of $k$.
In Figs. 1a-c we have plotted the squeezing factors $S_{j}(t)$ given by (11)
after substituting from (19) against time $t$ for
shown values of the parameters.
Further, in these figures first quadrature is always represented by the solid
curve, whereas second quadrature is represented by the dashed curve.
Now apart from the case $\phi=\frac{\pi}{4}$ which will be discussed
shortly, one can observe that at $t=0$ there is squeezing in the $K_{x}$
quadrature as expected since PCS are a type of squeezed states
depending on the value of $\phi$. When the
time increases exchange of energy between modes starts to play a
role, and consequently squeezing transfers to the second quadrature,
and in the first one it disappears. This behaviour
is periodically repeated as the interaction time increases.
Further, it is clear that larger the
parameter $|\xi|$, greater the squeezing which can be obtained.
It is important  mentioning
that squeezing can be realized even if the initial states are not
squeezed. This fact is demonstrated for the case $\phi=\frac{\pi}{4}$ where
PCS are not squeezed (this is clear from
Fig. 1c  at $t=0$), however, at later times
periodical squeezing is generated which  can be switched between the two
 quadratures. As we have shown before such behaviour
 can   periodically appear  with period $4\pi/g$.
Indeed such behaviours require that
$\lambda^{2}_{3}>\lambda^{2}_{1}+\lambda^{2}_{2}$, otherwise the initial
squeezing of PCS  will  vanish when the interaction time increases
since the solutions (17) in this case include  hyperbolic functions
which are monotonically increasing.

We proceed by focusing the attention on the behaviour of BGCS \cite{baru}
specified by (7). The required quantities to discuss $SU(1,1)$-squeezing
related to this state are

${\displaystyle \langle (\Delta K_{x}(t))^{2}\rangle_{\rm b} =
2|f(t,-1)|^{2} \left[n+ \frac{|z| I_{2n}(2|z|)}{I_{2n-1}(2|z|)}\right]-
S^{(+)}(t)[z^{*}f(t,-1)+z f^{*}(t,-1)]
} \hfill $

${\displaystyle + |z| S^{(+)2}(t)\left[ |z|\left(1-
\frac{ I^{2}_{2n}(2|z|)}{I^{2}_{2n-1}(2|z|)}\right)+(1-2n)
\frac{ I_{2n}(2|z|)}{I_{2n-1}(2|z|)}\right]}, \hfill (21a) $

${\displaystyle \langle (\Delta K_{y}(t))^{2}\rangle_{\rm b} =
2|g(t,-1)|^{2}\left[n+ \frac{|z| I_{2n}(2|z|)}{I_{2n-1}(2|z|)}\right]-
V^{(-)}(t)[z^{*}g(t,-1)+z g^{*}(t,-1)]
} \hfill $

${\displaystyle
+ |z| V^{(-)2}(t)\left[ |z|\left(1-
\frac{ I^{2}_{2n}(2|z|)}{I^{2}_{2n-1}(2|z|)}\right)+(1-2n)
\frac{ I_{2n}(2|z|)}{I_{2n-1}(2|z|)}\right]}, \hfill (21b) $

\noindent and

${\displaystyle \langle K_{z}(t)\rangle_{\rm b} =
 R_{3}(t,1) \left[n+ \frac{|z| I_{2n}(2|z|)}{I_{2n-1}(2|z|)}\right]
+ z^{*}h(t)+z h^{*}(t)}, \hfill (21c) $

\noindent where $f(t,-1), g(t,-1)$ and $h(t)$ are given in
(20).
As we mentioned earlier BGCS is similar to the Glauber coherent state,
i.e. it is a minimum-uncertainty state. However, it has been shown  that the
superposition of such states (even- and odd-BGCS)  can produce squeezing
as a result of the quantum mechanical interference between the
components of the state in phase space \cite{bann}. Also in the model under
discussion this state can  evolve to produce squeezing  (see Fig. 2 for shown
values of the parameters). It is clear that squeezing is generated and
interchanged between the two components provided that $
\lambda^{2}_{3}>\lambda^{2} _{1}+\lambda^{2} _{2}$.

\begin{figure}[h]%
   \includegraphics[width=8cm]{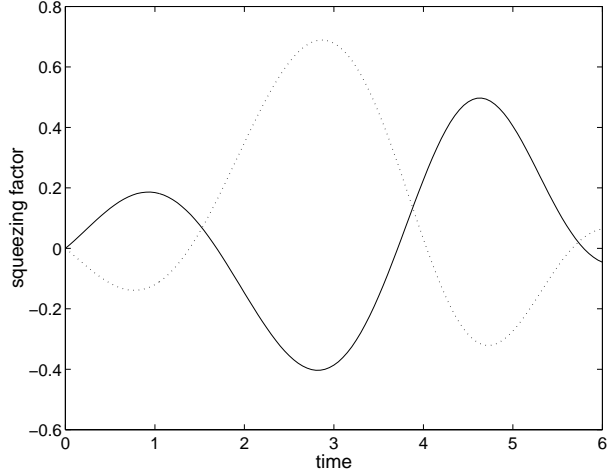}
    \caption{
Squeezing factor $S_{j}(t)$ of  BGCS against time $t$ for $n
=2,z=10\exp (i\pi)$ and $(\lambda_{1},\lambda _{2},\lambda
_{3})=(0.1,0.25,1)$ : first quadrature (solid curve), second
quadrature (dot curve). }
  \label{fig2}
\end{figure}

Second, we  study the $SU(2)$-squeezing in terms of $SU(2)$ CS (8)  as we
did before.  After straightforward
 calculations  the quadrature variances $\langle (\Delta
K_{x}(t))^{2}\rangle_{\rm u2}$ and
$\langle (\Delta K_{y}(t))^{2}\rangle_{\rm u2} $ as well as $\langle
 K_{z}(t)\rangle_{\rm u2}$ are

${\displaystyle \langle (\Delta K_{x}(t))^{2}\rangle_{\rm u2} =
2j\left\{ \frac{[S^{(+)}(t)-\mu^{*}f(t,1)-\mu f^{*}(t,1)][\mu^{*}f(t,1)
+\mu f^{*}(t,1)]}{(1+|\mu|^{2})}\right. } \hfill $

${\displaystyle
+|f(t,1)|^{2}  +\left.\frac{|\mu|^{2}[S^{(+)}(t)-\mu^{*}f(t,1)-\mu
f^{*}(t,1)]^{2}}{(1+|\mu|^{2})^{2}}\right\}}, \hfill (22a) $

${\displaystyle \langle (\Delta K_{y}(t))^{2}\rangle_{\rm u2} =
2j\left\{\frac{[V^{(+)}(t)-\mu^{*}g(t,1)-\mu g^{*}(t,1)][\mu^{*}g(t,1)
+\mu g^{*}(t,1)]}{(1+|\mu|^{2})}\right. } \hfill $

${\displaystyle
+|g(t,1)|^{2}+\left.\frac{|\mu|^{2}[V^{(+)}(t)-\mu^{*}g(t,1)-\mu
g^{*}(t,1)]^{2}}{(1+|\mu|^{2})^{2}}\right\}}, \hfill (22b) $

\noindent and

${\displaystyle \langle K_{z}(t)\rangle_{\rm u2} =
\frac{2j}{(1+|\mu|^{2})} \left\{ R_{3}(t,1) (|\mu|^{2}-1)
+2[\mu^{*}h(t)+\mu h^{*}(t)]\right\}}, \hfill (22c) $
\noindent where $f(t,1), g(t,1)$ and $h(t)$ are given in
(20).

\begin{figure}[h]%
    \includegraphics[width=8cm]{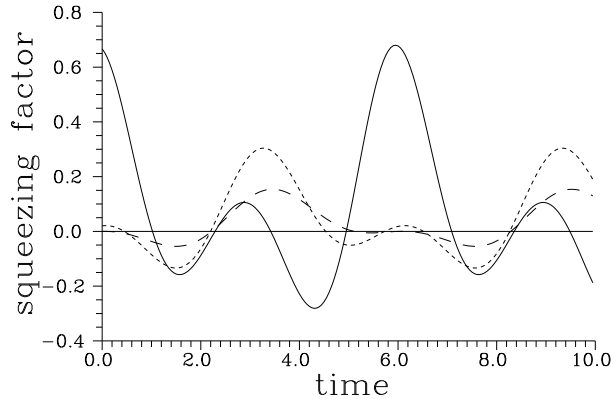}
    \caption{
Squeezing factor $S_{1}(t)$  (first quadrature) of
$SU(2)$-squeezing against time $t$ for $(\lambda_{1},\lambda
_{2},\lambda _{3})=(0.1,0.25,1)$, $ \phi =\frac{\pi }{2} $ and
$|\mu |=0.5$ (solid curve),$10$ (long-dashed curve), $100$
(short-dashed curve). Straight-line is the squeezing bound. }
  \label{fig3}
\end{figure}

In Fig. 3 we have plotted squeezing factor of the first quadrature against
time $t$ for the shown values of the parameters. From this figure
one can observe that there is no initial squeezing and this is in contrast
with  $SU(1,1)$-squeezing case (compare solid curves in Figs. 1 and Fig. 3).
As a result of the interaction of the field with the
material media squeezing can occur periodically with  maximum value smaller
than  for  $SU(1,1)$-squeezing.
Also one may observe that the degree of squeezing decreases as
 the values of $|\mu|$ increase (i.e. decreasing the initial mean photon
 number)  and this is in contrast with
$SU(1,1)$-squeezing where the opposite situation is established for
a given $|\xi|$ (as it is well known that the initial mean photon number
increases as $|\xi|$ increases).
In other words, when the initial mean photon number increases
the degrees of squeezing of both $SU(1,1)$- and $SU(2)$-squeezing
increase, too.

\section{Conclusions and remarks}

In this work we have studied
 $SU(1,1)$- and $SU(2)$-squeezing  of  interacting systems of radiation
modes in a quadratic medium in the framework of Lie algebra.
Particular examples have been given for  three mode case, however, the
model is quite general and may be applied to any  Hamiltonian
consisting of a set of operators obeying these kinds of Lie algebra.
In other words, from Hamiltonian (15) one can recognize that the boson
operators are not explicitly involved and the models become
indistinguishable.
This means that  if we have a model (say) which includes several
modes,
but its Hamiltonian can be represented as a linear combination of
$SU(1,1)$- or $SU(2)$-squeezing Lie
algebra generators, the behaviour of the degree of squeezing
of this model can be the same as discussed here.
For the considered models we have shown that squeezing is reached
for both PCS, BGCS and
$SU(2)$ CS, and can be periodically recovered provided  that $g$ is real.

We conclude this article by referring to \cite{hil} where two kinds of
two-mode squeezing (sum and difference squeezing) have been discussed.
Sum squeezing is described by operators which form a representation of the
$SU(1,1)$ Lie algebra, whereas operators of difference squeezing form
$SU(2)$ Lie algebra. Both of these kinds can be turned into normal
squeezing and consequently can be detected. Unfortunately,
this situation cannot be established here, where the Hamiltonian itself is
represented in terms of the quadrature operators and  any
 modification in the quadratures should be reflected in the structure of the
Hamiltonian. More illustratively, the used quadratures in this article
are represented bilinearly in bosonic operators and consequently they
can be converted into normal
squeezing. That is restricting ourselves on $SU(1,1)$ squeezing and considering
modes 1 and 2 are strong, they can be replaced by $|\Gamma_{j}|
\exp(i\phi_{j}), j=1,2$ where $|\Gamma_{j}|$ and $\phi_{j}$ are
their amplitudes
and  phases, further  taking  $\phi_{2}=\phi_{1}+\pi/2$. In this case the
quadratures (13) reduce to those of normal squeezing as

${\displaystyle
L_{x} =-|\Gamma_{2}|[\hat{A}_{3}\exp(-i\phi_{1})+
\hat{A}^{\dagger}_{3}\exp(i\phi_{1}) ],
\quad
L_{y} =i|\Gamma_{1}|[\hat{A}_{3}\exp(-i\phi_{1})-
\hat{A}^{\dagger}_{3}\exp(i\phi_{1}) ],} \hfill $

${\displaystyle
L_{z} =-2|\Gamma_{1}||\Gamma_{2}|
}. \hfill (23) $

\noindent However, the price is payed that the Hamiltonian becomes
a linear
combination of creation and annihilation operators which cannot provide
 squeezing as well as the rules of the Lie algebra are not established.
In conclusion, we have showed that
special types of three modes interacting bilinearly  in a nonlinear crystal can
provide squeezing.
This can be achieved in sum (difference)-frequency generation where
the interaction arises from the second-order polarizability of the nonlinear
medium. Squeezing in the quadratures $K_{j}, j=x,y$ of the input field
can be observed by studying the standard quadrature of the output field
\cite{aga}, e.g. through heterodyne detector.
Moreover, such realization seems to be more feasible using
the $SU(2)$ and $SU(1,1)$ interferometers \cite{bern}.
In the case of  $SU(1,1)$ interferometer the beam splitters of
a conventional interferometer have been replaced by the four-wave mixers
and consequently it has a simpler construction than
the $SU(2)$ interferometer. Indeed, this fact
together with periodic solution of equations of motion
 with the period $4\pi/g$
 can  be used for obtaining squeezing on a rather long time scale.

{\bf Acknowledgments}

 J. P. and F. A. A. E-O. acknowledge the
partial support from the Project VS96028 and Research Project CEZ:
J 14/98 of Czech Ministry of Education and from the Project
202/00/0142 of Czech Grant Agency. One of us (M. S. A.) is
grateful for the financial support from the Project Math 1418/19
of the Research Centre, College of Science, King Saud University.

\end{document}